# Dual-band nonreciprocal thermal radiation by coupling optical Tamm states in magnetophotonic multilayers

Jun Wu[1,*], Feng Wu[2], Tiancheng Zhao[3], Mauro Antezza[4,5], and Xiaohu Wu[6,*]

1. Key Laboratory of Advanced Perception and Intelligent Control of High-end Equipment, Ministry of Education, College of Electrical Engineering, Anhui Polytechnic University, Wuhu, 241000, China

2. School of Optoelectronic Engineering, Guangdong Polytechnic Normal University, Guangzhou 510665, China

3. Beijing Aerospace Institute for Metrology and Measurement Technology, Beijing 100076, China

4. Laboratoire Charles Coulomb (L2C), UMR 5221 CNRS-Université de Montpellier, F-34095 Montpellier, France

5. Institut Universitaire de France, 1 rue Descartes, F-75231 Paris Cedex 05, France

6. Shandong Institute of Advanced Technology, Jinan 250100, China

*Corresponding author: mailswj2011@163.com, xiaohu.wu@iat.cn

**Abstract:** Kirchhoff's law is one of the most fundamental law in thermal radiation. The violation of traditional Kirchhoff's law provides opportunities for higher energy conversion efficiency. Various micro-structures have been proposed to realize single-band nonreciprocal thermal emitters. However, dual-band nonreciprocal thermal emitters remain barely investigated. In this paper, we introduce magneto-optical material into a cascading one-dimensional (1-D) magnetophotonic crystal (MPC) heterostructure composed of two 1-D MPCs and a metal layer. Assisted by the nonreciprocity of the magneto-optical material and the coupling effect of two optical Tamm states (OTSs), a dual-band nonreciprocal lithography-free thermal emitter is achieved. The emitter exhibits strong dual-band nonreciprocal radiation at the wavelengths of 15.337 μm and 15.731 μm when the external magnetic field is 3 T and the angle of incidence is 56°. Besides, the magnetic field distribution is also calculated to confirm that the dual-band nonreciprocal radiation originates from the coupling effect between two OTSs. Our work may pave the way for constructing dual-band and multi-band nonreciprocal thermal emitters.

# 1. Introduction

Kirchhoff's law is one of the most fundamental law in thermal radiation, thus it has long been a subject of great interest [1]. The traditional Kirchhoff's law of thermal radiation states the directional spectrum emissivity of an object should be equal to its directional spectrum absorptivity, which provides a simple way to calculate the emissivity via absorptivity [2-7]. However, some researchers have pointed out that the traditional Kirchhoff's law is only appliable for reciprocal materials, rather than nonreciprocal materials, which does not obey the Lorentz reciprocity [8, 9]. More recently, much attention has been paid to the generalized Kirchhoff's law, which is appliable for both reciprocal and nonreciprocal materials [10-12]. In 2012, Green has proved that the conversion efficiency limit of nonreciprocal thermal emitters could exceed the classic Shockley-Queisser (SQ) limit [13]. Therefore, the design of nonreciprocal thermal emitters is of great importance.

Recently, various nonreciprocal materials, such as magneto-optical materials and Weyl semimetals, were utilized to break the traditional Kirchhoff's law and construct nonreciprocal thermal emitters [14-25]. For instance, the nonreciprocal radiation near wavelength of 16 μm is realized with magneto-optical material InAs under an external magnetic field of 3 T [14]. In addition, it is found that Weyl semimetals showed stronger nonreciprocity at mid-infrared wavelengths without external magnetic fields [15-17, 21]. As far as we know, previous studies are mainly focused on single-band nonreciprocal thermal emitters [14, 15, 17-20]. Nevertheless, dual-band and multi-band nonreciprocal thermal emitters are more worthy of intense interests since they possess promising applications in renewable energy systems. Up until now, dual-band nonreciprocal thermal emitters have remained barely investigated. Very recently, researches utilized four-part periodic metal gratings to realize dual-band nonreciprocal emitters [22]. However, the fabrication of four-part periodic metal grating needs lithography technique, which considerably increases the complexity and the cost of the fabrication process [26,27]. A question arises naturally: Can we achieve dual-band nonreciprocal emitters based on lithography-free one-dimensional (1-D) structures?

In 2005, Kavokin et al. discovered a kind of special localized optical states called optical Tamm states (OTSs) in lithography-free 1-D heterostructures composed of two kinds of 1-D photonic crystals (1-D PCs) [28]. It is well-known that 1-D PCs can be viewed as optical mirrors within their photonic bandgaps (PBGs) [29, 30]. Hence, as the reflection phase matching condition is satisfied, 1-D heterostrucrures composed of two kinds of 1-D PCs can be viewed as a Fabry-Perot cavity, leading to OTSs [31, 32]. Then, in 2007, Kaliteevski et al. found that OTSs can also be realized in 1-D heterostructures composed of a 1-D PC and a metal layer since the metal layer can also be viewed as an optical mirror below its plasma frequency [33]. Owing to the unique resonant property, OTSs have been widely employed to greatly enhance optical absorption [34-37], Faraday rotation [38-40], and optical nonlinear effect [41]. Over the past decade, researchers have cascaded several 1-D heterostructures together and studied the coupling effect between different OTSs [42-44], which provides us an opportunity to realize dual-band nonreciprocal emitters based on lithography-free 1-D heterostructures.

In this work, we introduce a kind of magneto-optical materials InAs into a 1-D PC heterostructure composed of two kinds of 1-D PCs to construct a 1-D magnetophotonic crystal (MPC) heterostructure. Then, we put a metal layer (Al) behind the 1-D MPC heterostructure. Therefore, the OTS in 1-D MPC heterostructure composed of two kinds of 1-D MPCs can couple with the OTS in 1-D heterostructure composed of a 1-D MPC and a metal layer, leading to dual OTSs. Assisted by the dual OTSs and the nonreciprocity of the magneto-optical material, we achieve dual-band nonreciprocal thermal radiation in a 1-D MPC heterostructure. Then, the magnetic field intensity distributions are calculated to confirm the physical mechanism behind such phenomenon. Finally, the influence of the geometrical parameters on the dual-band nonreciprocal radiation is investigated.

## 2. Model and theoretical method

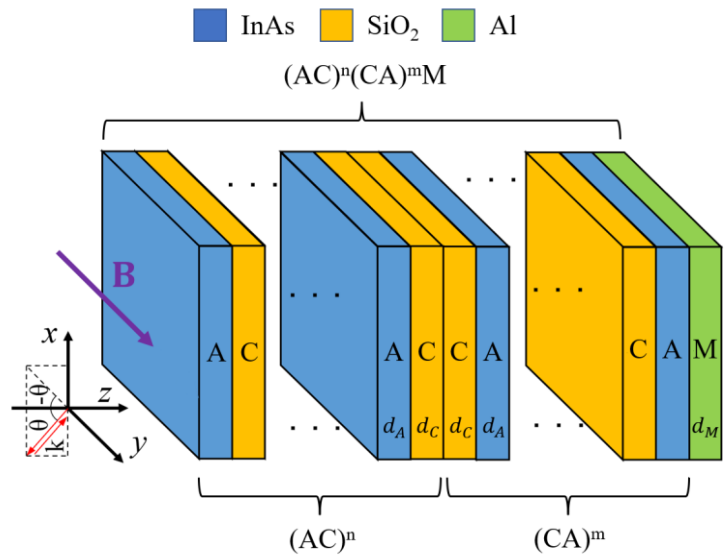

Fig. 1 Schematic of the proposed dual-band nonreciprocal thermal emitter. The external magnetic field B is along *y*-axis.

The proposed dual-band nonreciprocal thermal emitter is shown in Fig. 1, which consists of a 1-D PC $(AC)^n$, a 1-D PC $(CA)^m$ and a metal layer M. The whole structure $(AC)^n(CA)^m M$ can be viewed as the combination of the following two heterostructures. The first heterostructure is $(AC)^n(CA)^m$, which is composed of two kinds of 1-D PCs while the second heterostructure is $(CA)^m M$, which composed of a 1-D PC and a metal layer. These two heterostructures support two coupled OTSs, which leads to dual-band nonreciprocal radiation.

In this work, the material of medium A is chosen to be a kind of magneto-optical materials InAs. When an external magnetic field with intensity B is applied along the $y$-axis, the relative permittivity tensor of InAs can be expressed in [9, 18]:

$$\varepsilon = \begin{bmatrix} \varepsilon_{xx} & 0 & \varepsilon_{xz} \\ 0 & \varepsilon_{yy} & 0 \\ \varepsilon_{zx} & 0 & \varepsilon_{zz} \end{bmatrix}, \tag{1}$$

where

$$\varepsilon_{xx} = \varepsilon_{zz} = \varepsilon_\infty - \frac{\omega_p^2(\omega+i\Gamma)}{\omega\left[(\omega+i\Gamma)^2 - \omega_c^2\right]}, \tag{2}$$

$$\varepsilon_{xz} = -\varepsilon_{zx} = i\frac{\omega_p^2 \omega_c}{\omega\left[(\omega+i\Gamma)^2 - \omega_c^2\right]}, \tag{3}$$

$$\varepsilon_{yy} = \varepsilon_\infty - \frac{\omega_p^2}{\omega(\omega+i\Gamma)}. \tag{4}$$

Here, the detailed definitions and parameter values shown in Eqs. (2)-(4) can be found in Ref. [18]. The material of the medium C is chosen to be silica (SiO$_2$) with a refractive index of 1.45 [45]. The material of the metal M is chosen to be aluminum (Al). The relative permittivity of Al can be described by the Drude model [46]

$$\varepsilon_{Al} = \varepsilon_{\infty} - \frac{\omega_p^2}{\omega^2 + j\omega\Gamma}, \tag{5}$$

where $\varepsilon_\infty$=1, $\Gamma$=1.24×10$^{14}$ rad/s and $\omega_p$=2.24×10$^{16}$ rad/s [46].

When a TM (transverse magnetic, with the magnetic field along the direction of *y*-axis)-polarized plane wave is incident (in Fig. 1 from the left side) with an angle $\theta$, the spectral directional absorption and emission of the structure can be calculated by [11]

$$\begin{array}{l} \alpha(\theta,\lambda)=1-R(\theta,\lambda)-T(\theta,\lambda), \alpha(-\theta,\lambda)=1-R(-\theta,\lambda)-T(-\theta,\lambda)\\ e(\theta,\lambda)=1-R(-\theta,\lambda)-T(\theta,\lambda), e(-\theta,\lambda)=1-R(\theta,\lambda)-T(-\theta,\lambda) \end{array}. \tag{6}$$

Here, $R(\theta, \lambda)$ and $T(\theta, \lambda)$ are the reflection and transmission spectra for incident angle of $\theta$, respectively, while $R(-\theta, \lambda)$ and $T(-\theta, \lambda)$ denote the corresponding reflection and transmission spectra under the incident angle of $-\theta$, respectively. The difference between emission and absorption is defined as $\eta=|\alpha-e|$, measuring the nonreciprocal radiation. All of which can be obtained through the transfer matrix method (TMM) [47-50].

To introduce the TMM in detail, we first take one slab with thickness *d* as an example. The relative permittivity tensor of the slab is the same as Eq. (1). It is noted that the off-diagonal elements are zero for SiO$_2$ and Al layers. The electromagnetic fields in the slab can be expressed as

$$\mathbf{H} = \mathbf{U}(z)\exp(j\omega t - jk_x x), \text{where } \mathbf{U}=(0,U_y,0). \tag{7}$$

$$\mathbf{E} = j(\mu_0/\varepsilon_0)^{1/2}\mathbf{S}(z)\exp(j\omega t - jk_x x), \text{where } \mathbf{S}=(S_x,0,S_z). \tag{8}$$

Note that $\varepsilon_0$ and $\mu_0$ are the absolute permittivity and permeability of vacuum, respectively, $\omega$ is the angular frequency, $k_x=k_0\sin\theta$, where $k_0$ is the wavevector in the vacuum. Substituting Eqs. (7) and (8) into the Maxwell's equations and setting $K_x = k_x/k_0$, we obtain the following differential equations:

$$\frac{\partial}{\partial z}\begin{pmatrix} U_y \\ S_x \end{pmatrix} = k_0 \begin{bmatrix} \varepsilon_a/\varepsilon_d K_x & \varepsilon_d - \varepsilon_a^2/\varepsilon_d \\ K_x^2/\varepsilon_d - 1 & -\varepsilon_a/\varepsilon_d K_x \end{bmatrix} \begin{pmatrix} U_y \\ S_x \end{pmatrix}. \tag{9}$$

The electromagnetic fields in the slab can be described by the eigenvalues and eigenvectors of the coefficient matrix:

$$U_y(z) = w_{11} c^+ \exp(k_0 q^+ z) + w_{12} c^- \exp(k_0 q^- (z-d)), \tag{10}$$

$$S_x(z) = w_{21} c^+ \exp(k_0 q^+ z) + w_{22} c^- \exp(k_0 q^- (z-d)), \tag{11}$$

where $w_{im}$ is the corresponding element of the eigenvector matrix $\mathbf{W}$ of the coefficient matrix in Eq. (9), $q^+$ and $q^-$ are the two eigenvalues of matrix $\mathbf{A}$ with negative real part and positive real part, respectively, $c^+$ and $c^-$ are unknowns. By applying the boundary conditions, i.e., the tangential components of the magnetic and electric field vectors should be continuous at the interface, it can be shown that

$$\begin{pmatrix} 1 \\ -j\sin\theta \end{pmatrix} + \begin{pmatrix} 1 \\ j\sin\theta \end{pmatrix} r = \mathbf{W} \begin{pmatrix} 1 & 0 \\ 0 & \exp(-k_0 q^- d) \end{pmatrix} \begin{pmatrix} c^+ \\ c^- \end{pmatrix}, \tag{12}$$

$$\mathbf{W} \begin{pmatrix} 1 & 0 \\ 0 & \exp(k_0 q^+ d) \end{pmatrix} \begin{pmatrix} c^+ \\ c^- \end{pmatrix} = \begin{pmatrix} 1 \\ -j\sin\theta \end{pmatrix} t, \tag{13}$$

where $r$ and $t$ are the reflection and transmission coefficients, respectively. From Eqs. (12) and (13), the reflection and transmission coefficients of one slab can be obtained.

We can apply the above analysis to arbitrary $L$-layer structure. All the boundary conditions are

$$\begin{pmatrix} 1 \\ -j\sin\theta \end{pmatrix} + \begin{pmatrix} 1 \\ j\sin\theta \end{pmatrix} r = \mathbf{W}_{(1)} \begin{pmatrix} 1 & 0 \\ 0 & X_{(1)} \end{pmatrix} \begin{pmatrix} c^+_{(1)} \\ c^-_{(1)} \end{pmatrix}, \tag{14}$$

$$\mathbf{W}_{(l-1)}\begin{pmatrix} 1 & 0 \\ 0 & Y_{(l-1)} \end{pmatrix}\begin{pmatrix} c^+_{(l-1)} \\ c^-_{(l-1)} \end{pmatrix} = \mathbf{W}_{(l)}\begin{pmatrix} 1 & 0 \\ 0 & X_{(l)} \end{pmatrix}\begin{pmatrix} c^+_{(l)} \\ c^-_{(l)} \end{pmatrix}, \quad (15)$$

$$\mathbf{W}_{(L)}\begin{pmatrix} 1 & 0 \\ 0 & Y_{(L)} \end{pmatrix}\begin{pmatrix} c^+_{(L)} \\ c^-_{(L)} \end{pmatrix} = \begin{pmatrix} 1 \\ -j\sin\theta \end{pmatrix} t. \quad (16)$$

where $l=2,3,\ldots,L$. $\mathbf{W}_{(l)}$ is the eigenvector matrix for the $l^{th}$ slab. $X_{(l)}$ and $Y_{(l)}$ are defined as $\exp(-k_0 q^-_{(l)} d_{(l)})$ and $\exp(k_0 q^+_{(l)} d_{(l)})$, respectively. $d_{(l)}$ is the thickness of the $l^{th}$ slab. $q^+_{(l)}$ and $q^-_{(l)}$ are eigenvalues in the $l^{th}$ slab with negative real part and positive real part, respectively. $c^+_{(l)}$ and $c^-_{(l)}$ are the unknowns. By solving Eqs. (14-16), the reflection and transmission coefficients of the $L$-layer structure can be obtained. Finally, the absorption and emission can be obtained.

## 3. Results and discussion

The external magnetic field is fixed as 3 T in this work, which is experimentally achievable [9]. The numbers of the unit cells in 1-D PCs are chosen to be $n=4$ and $m=4$. To realize high emission, the thickness parameters and angle of incidence are optimized as: $d_A$=4.6 μm, $d_C$=3.23 μm, $d_M$=0.2 μm, and $\theta$=56°. It is noted that the angle of incidence is typically larger than 45° to realize strong nonreciprocal radiation [9, 21].

In Fig. 2(a), we show the calculated absorption and emission spectra at the angle of incidence of 56° with B=0 and B=3 T, respectively. One can see that the absorption and emission spectra are overlapped when B=0. In addition, two high absorption (emission) peaks (0.954 and 0.998) are achieved. These two peaks originate from the coupling effect between two OTSs, which is confirmed by the magnetic field distribution as we discussed latter. When the strength of the external magnetic field B increases from 0 to 3 T, both absorption peaks and emission peaks exhibit blueshift. Most importantly, the absorption and the emission spectra do not overlap with each other, indicating the violation of the traditional Kirchhoff's law. Fig. 2(b) gives the difference between absorption and emission $\eta = |\alpha - e|$ as a function of wavelength when B=3 T. The difference at the wavelengths of 15.337 μm and 15.731 μm can reach 0.86 and 0.92, respectively, both of which exhibit strong nonreciprocal radiation. Such phenomenon demonstrates the dual-band strong nonreciprocal radiation property of the proposed emitter.

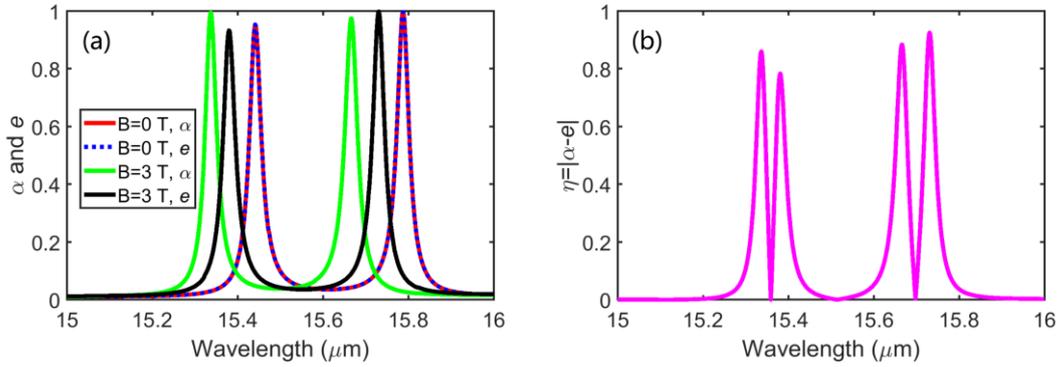

Fig. 2 (a) Absorption ($\alpha$) and emission ($e$) spectra at $\theta$=56º for B=0 and B=3 T. (b) Difference between the absorption and emission $\eta = |\alpha - e|$ as a function of wavelength at $\theta$=56º.

To confirm that the dual-band nonreciprocal radiation originates from the coupling effect between two OTSs, we calculate the magnetic field distributions ($|H_y|$) at two resonant wavelengths 15.337 μm and 15.731 μm in Figs. 3(a) and 3(b), respectively. The red and blue solid lines represent the cases at $\theta$=56º and $\theta$=-56º, respectively. As shown in Fig. 3, both the magnetic field at two resonant wavelengths 15.337 μm and 15.731 μm are greatly enhanced around two interfaces. The first interface (shown by the green dashed line) is located between the 1-D PC (AC)$^4$ and the 1-D PC (CA)$^4$ while the second interface (shown by the black dashed line) is located between the 1-D PC (CA)$^4$ and the metal layer M. The calculated magnetic field distribution confirms that the dual-band nonreciprocal radiation originates from the coupling effect between two OTSs. As shown in Fig. 3(a), one can see that the magnetic field at $\theta=56^o$ are stronger than that at $\theta=-56^o$. At the interface between the air and the first slab of the proposed emitter, the strength of magnetic field is close to 1 at $\theta=56^o$, while it is close to 2 at $\theta=-56^o$, indicating that strong absorption occurs at $\theta=56^o$ while strong reflection occurs at $\theta=-56^o$. The situation is quite similar in Fig. 3(b). The only difference is that the magnetic field at $\theta=-56^o$ are stronger than that at $\theta=56^o$, which can be explained as follows. In Fig. 3(a), the wavelength (15.337 μm) is selected at the absorption peak in the first nonreciprocal band. However, in Fig. 3(b), the wavelength (15.731 μm) is selected at the emission peak in the second nonreciprocal band.

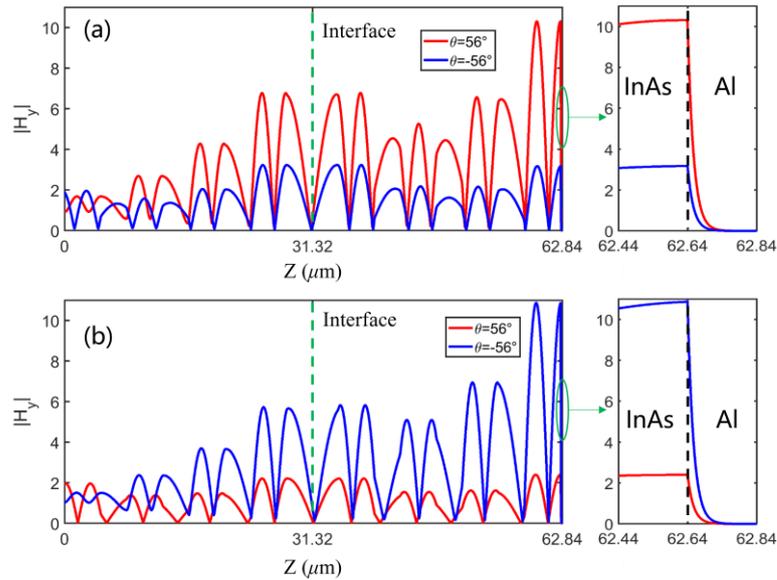

Fig. 3 Magnetic field distribution ($|H_y|$) of the proposed emitter (AC)$^4$(CA)$^4$M at the wavelengths (a) 15.337 μm and (b) 15.731 μm at $\theta=56°$ and $\theta=-56°$.

To further prove the coupling effect between two OTSs, two uncoupled OTSs should be shown in the heterostructure $(AC)^4(CA)^4$ and the heterostructure $(CA)^4M$. First, we show the uncoupled OTS in the heterostructure $(AC)^4(CA)^4$. Fig. 4(a) gives the absorption spectrum of the heterostructure when B=3 T and $\theta=56^o$. One can see that an absorption peak is located around the wavelength of 15.505 μm, which indicates that an OTS occurs [34]. Here the peak absorption is not high since the loss of InAs is not large. Fig. 4(b) gives the corresponding magnetic field distributions ($|H_y|$) at the OTS wavelength 15.505 μm. One can see that at the OTS wavelength, the magnetic field is strongly localized around the interface between two 1DPCs.

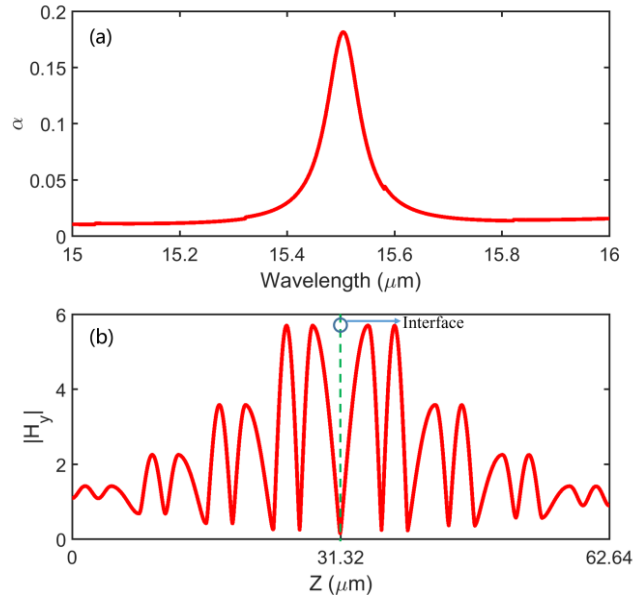

Fig. 4 (a) Absorption spectrum of the heterostructure $(AC)^4(CA)^4$ when B=3 T and $\theta=56^o$. (b) Magnetic field distribution ($|H_y|$) of the heterostructure $(AC)^4(CA)^4$ at the wavelength of 15.505 μm.

Then, we show another uncoupled OTS in the heterostructure $(CA)^4M$. Fig. 5(a) gives the absorption spectrum of the heterostructure when B=3 T and $\theta=56^o$. One can see that an absorption peak is located around the wavelength of 15.493 μm, which indicates that an OTS occurs [34]. Fig. 5(b) gives the corresponding magnetic field distributions ($|H_y|$) at the OTS wavelength 15.493 μm. One can see that at the OTS wavelength, the magnetic field is strongly localized around the interface between the 1DPC and the Al layer. By combining the heterostructure $(AC)^4(CA)^4$ with and the heterostructure $(CA)^4M$, two OTSs couple with each other, leading to dual-band nonreciprocal radiation.

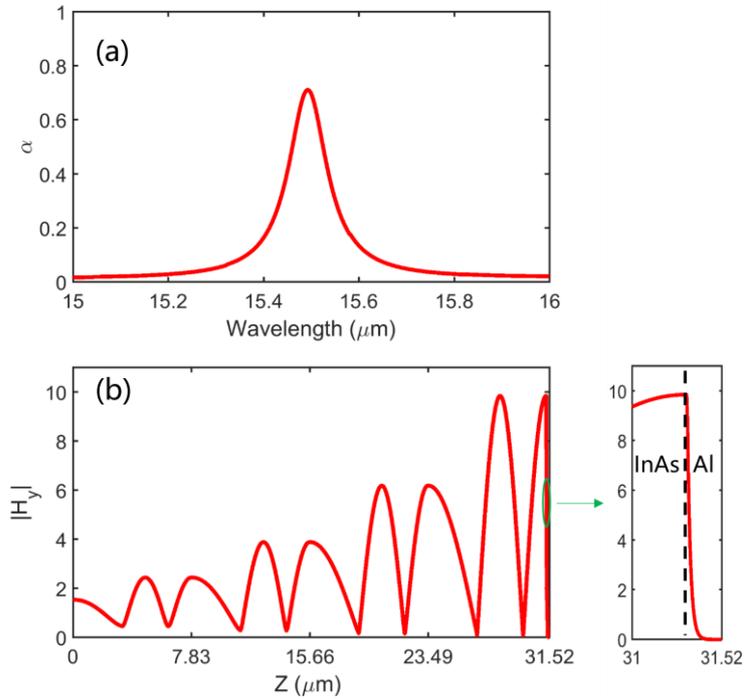

Fig. 5 (a) Absorption spectrum of the heterostructure $(CA)^4M$ at $\theta=56º$ for B=3 T. (b) Magnetic field distribution ($|H_y|$) of the heterostructure $(CA)^4M$ at the wavelength of 15.493 μm.

Finally, we investigate the influence of the periodic number and the layer thickness on the nonreciprocal radiation. Figs. 6(a)-6(d) give the absorption (*α*), emission (*e*) and difference between absorption and emission ($\eta=|\alpha-e|$) as a function of wavelength for different values of *n* and *m*. As shown in Fig. 6(a), the absorption and emission are smaller than 0.8 when *n*=*m*=3. Besides, the full widths at half maximum (FWHM) of the absorption and emission peaks are larger than those shown in Fig. 2(a). As a consequence, the difference between absorption and emission is smaller than 0.6. As shown in Fig. 6(b), the difference between absorption and emission at wavelengths of 15.44 and 15.6 μm reach 0.9 and 0.85 when $n=m=5$. Compared with Fig. 6(a) and Fig. 2(a), one can see that the absorption and emission peaks will become closer as the periodic numbers increase. As shown in Figs. 6(c) and 6(d), the results do not change to much when the periodic numbers are different ($n \neq m$). Compared Figs. 6(a)-6(d) with Fig. 2(a), it is clear that the performance is the best when $n=m=4$.

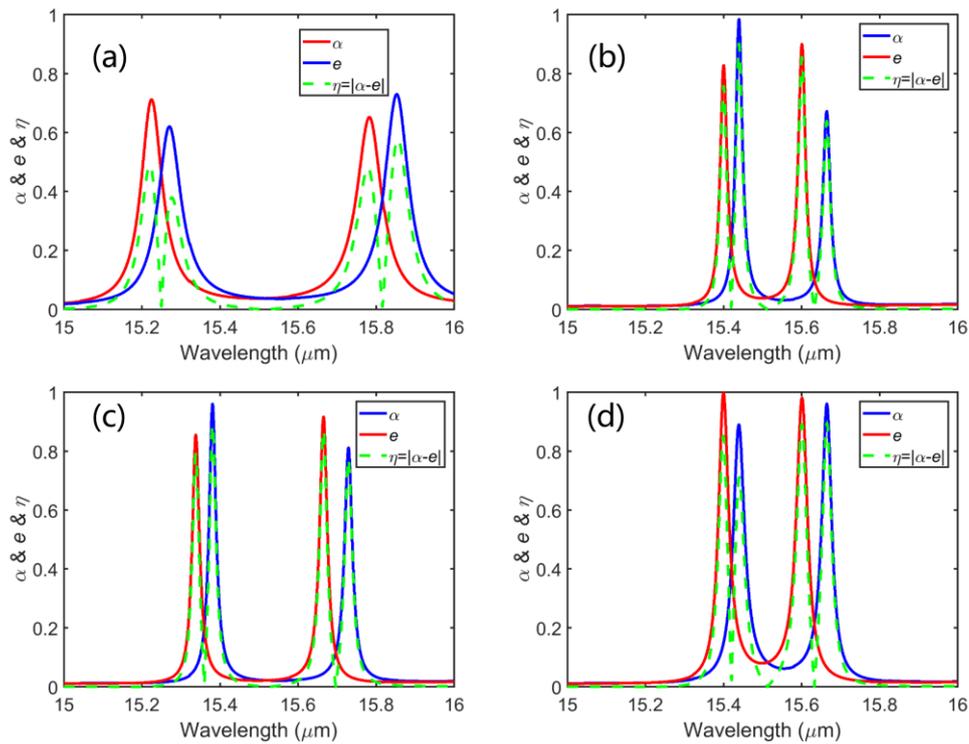

Fig. 6 Absorption, emission and difference between them as a function of wavelength with different values of $n$ and $m$: (a) $n=m=3$, (b) $n=m=5$, (c) $n=5$, $m=4$, and (d) $n=4$, $m=5$.

Fig. 7(a), 7(b) and 7(c) give the absorption, emission, and difference between them varying with the wavelength and the thickness of $SiO_2$ layer $d_C$, respectively.

The periodic numbers are chosen to be $n=m=4$. The thicknesses of InAs and Al layers are chosen to be $d_A=4.6$ μm and $d_M=0.2$ μm, respectively. It is clear that dual-band near-perfect absorption and emission can be realized. In addition, the peaks of absorption and emission shift toward larger wavelengths as the thickness of $SiO_2$ layer increases. It is noted that the results for only wavelength ranging from 15 μm to 16 μm is given. When the wavelength is smaller than 15 μm or larger than 16 μm, it is also possible to realize dual-band nonreciprocal radiation by adjusting the thickness of the $SiO_2$ layer. When the thickness of the $SiO_2$ layer is smaller than 2 μm or larger than 4.5 μm, the absorption and emission are almost identical, thus the difference between them is close to zero, as shown in Fig. 7(c).

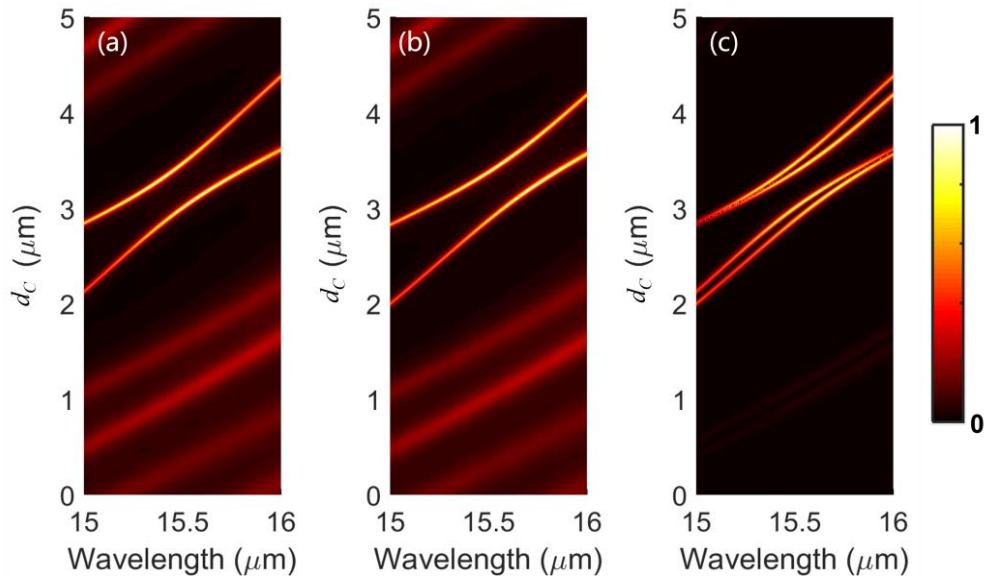

Fig. 7 (a) Absorption, (b) emission and (c) difference between them vary with the wavelength and the thickness of the $SiO_2$ layer $d_C$. The periodic numbers are chosen to be n=m=4. The thicknesses of InAs and Al layers are chosen to be $d_A=4.6$ μm and $d_M=0.2$ μm.

Fig. 8(a), 8(b) and 8(c) give the absorption, emission, and difference between them varying with the wavelength and the thickness of InAs layer $d_A$, respectively.

The periodic numbers are chosen to be n=m=4. The thicknesses of $SiO_2$ and Al layers are chosen to be $d_C$=3.23 μm and $d_M$=0.2 μm. As shown in Figs. 8(a) and 8(b), one can see that dual-band absorption and emission can be realized when the thickness of the InAs layer is smaller than 2 μm or larger than 4.1 μm. In addition, as shown in Fig. 8(c), dual-band strong nonreciprocal radiation can be achieved when the thickness of the InAs layer is smaller than 2 μm or larger than 4.1 μm. The dual-band nonreciprocal radiation shifts towards longer wavelengths as the thickness of the InAs layer increases, exhibiting the similar trend shown in Fig. 7(c). When the thickness of the InAs layer between 2 μm and 4.1 μm, the absorption and emission are almost identical, thus the difference between them is close to zero.

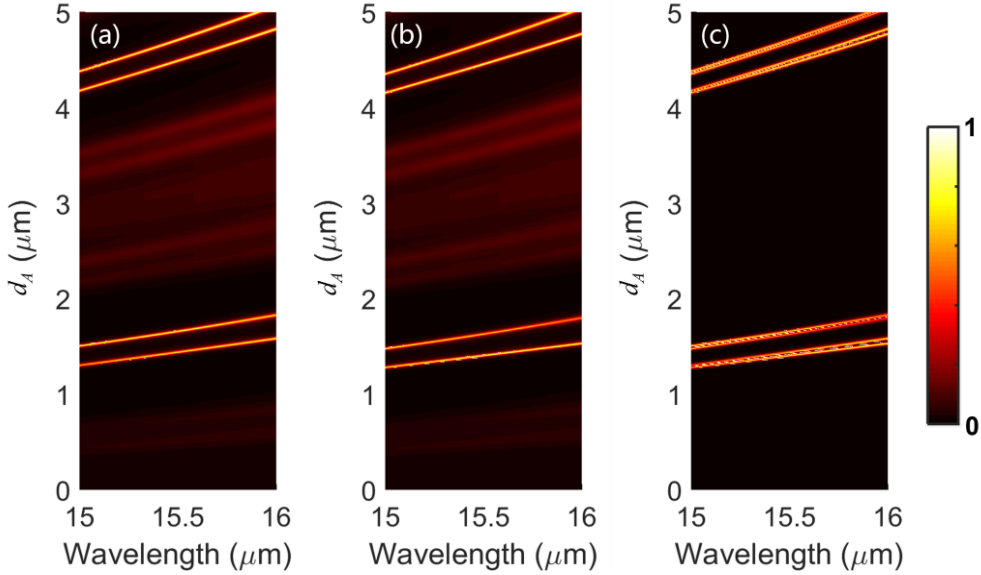

Fig. 8 (a) Absorption, (b) emission and (c) difference between them vary with the wavelength and the thickness of the InAs layer $d_A$. The periodic numbers are chosen to be n=m=4. The thicknesses of $SiO_2$ and Al layers are chosen to be $d_C$=3.23 μm and $d_M$=0.2 μm.

## 4. Conclusions

In summary, based on the coupling effect between two OTSs, strong dual-band nonreciprocal thermal radiation is realized in a cascading 1-D MPC heterostructure composed of two 1-D MPCs and a metal layer. The lithography-free emitter shows strong dual-band nonreciprocal radiation at the wavelength of 15.337 μm and 15.731 μm under an external magnetic field of 3 T when the incident angle is 56°. The coupling effect between two OTSs is confirmed by the magnetic field distribution. It is hoped that the method proposed in this work will promote the development of novel energy harvesting devices and nonreciprocal thermal emitters.

**Acknowledgements**

The authors acknowledge the support of the National Natural Science Foundation of China (Grant Nos. 61405217, 52106099 and 12104105)，the Zhejiang Provincial Natural Science Foundation (Grant No. LY20F050001), the Anhui Polytechnic University Research Startup Foundation (Grant No. 2020YQQ042), the Pre-research Project of National Natural Science Foundation of Anhui Polytechnic University (Grant No. Xjky02202003), the Natural Science Foundation of Shandong Province (Grant No. ZR2020LLZ004), and the Start-Up Funding of Guangdong Polytechnic Normal University (Grant No. 2021SDKYA033).